\documentstyle[12pt]{article}

\catcode`\@=11
\long\def\@makefntext#1{
\protect\noindent \hbox to 3.2pt {\hskip-.9pt
$^{{\ninerm\@thefnmark}}$\hfil}#1\hfill}		%CAN BE USED

 \def\@makefnmark{\hbox to 0pt{$^{\@thefnmark}$\hss}}  %ORIGINAL

\def\ps@myheadings{\let\@mkboth\@gobbletwo
\def\@oddhead{\hbox{}
\rightmark\hfil\ninerm\thepage}
\def\@oddfoot{}\def\@evenhead{\ninerm\thepage\hfil
\leftmark\hbox{}}\def\@evenfoot{}
\def\sectionmark##1{}\def\subsectionmark##1{}}

\newcounter{sectionc}\newcounter{subsectionc}\newcounter{subsubsectionc}
\renewcommand{\section}[1] {\vspace{0.6cm}\addtocounter{sectionc}{1}
\setcounter{subsectionc}{0}\setcounter{subsubsectionc}{0}\noindent
	{\bf\thesectionc. #1}\par\vspace{0.4cm}}
\renewcommand{\subsection}[1] {\vspace{0.6cm}\addtocounter{subsectionc}{1}
	\setcounter{subsubsectionc}{0}\noindent
	{\it\thesectionc.\thesubsectionc. #1}\par\vspace{0.4cm}}
\renewcommand{\subsubsection}[1]
{\vspace{0.6cm}\addtocounter{subsubsectionc}{1}
	\noindent {\rm\thesectionc.\thesubsectionc.\thesubsubsectionc.
	#1}\par\vspace{0.4cm}}

\newcounter{appendixc}
\newcounter{subappendixc}[appendixc]
\newcounter{subsubappendixc}[subappendixc]

\renewcommand{\appendix}[1] {\vspace{0.6cm}
        \refstepcounter{appendixc}
        \setcounter{figure}{0}
        \setcounter{table}{0}
        \setcounter{equation}{0}
        \renewcommand{\thefigure}{\Alph{appendixc}.\arabic{figure}}
        \renewcommand{\thetable}{\Alph{appendixc}.\arabic{table}}
        \renewcommand{\theappendixc}{\Alph{appendixc}}
        \renewcommand{\theequation}{\Alph{appendixc}.\arabic{equation}}
        \noindent{\bf Appendix \theappendixc #1}\par\vspace{0.4cm}}

\def\abstracts#1{{
	\centering{\begin{minipage}{30pc}\tenrm\baselineskip=12pt\noindent
	\centerline{\tenrm ABSTRACT}\vspace{0.3cm}
	\parindent=0pt #1
	\end{minipage}}\par}}

\renewenvironment{thebibliography}[1]
	{\begin{list}{\arabic{enumi}.}
	{\usecounter{enumi}\setlength{\parsep}{0pt}
\setlength{\leftmargin 1.25cm}{\rightmargin 0pt}
	 \setlength{\itemsep}{0pt} \settowidth
	{\labelwidth}{#1.}\sloppy}}{\end{list}}

\newcounter{itemlistc}
\newcounter{romanlistc}
\newcounter{alphlistc}
\newcounter{arabiclistc}

\newcommand{\fcaption}[1]{
        \refstepcounter{figure}
        \setbox\@tempboxa = \hbox{\tenrm Fig.~\thefigure. #1}
        \ifdim \wd\@tempboxa > 6in
           {\begin{center}
        \parbox{6in}{\tenrm\baselineskip=12pt Fig.~\thefigure. #1}
            \end{center}}
        \else
             {\begin{center}
             {\tenrm Fig.~\thefigure. #1}
              \end{center}}
        \fi}

\newcommand{\tcaption}[1]{
        \refstepcounter{table}
        \setbox\@tempboxa = \hbox{\tenrm Table~\thetable. #1}
        \ifdim \wd\@tempboxa > 6in
           {\begin{center}
        \parbox{6in}{\tenrm\baselineskip=12pt Table~\thetable. #1}
            \end{center}}
        \else
             {\begin{center}
             {\tenrm Table~\thetable. #1}
              \end{center}}
        \fi}

\def\@citex[#1]#2{\if@filesw\immediate\write\@auxout
	{\string\citation{#2}}\fi
\def\@citea{}\@cite{\@for\@citeb:=#2\do
	{\@citea\def\@citea{,}\@ifundefined
	{b@\@citeb}{{\bf ?}\@warning
	{Citation `\@citeb' on page \thepage \space undefined}}
	{\csname b@\@citeb\endcsname}}}{#1}}

\newif\if@cghi
\def\cite{\@cghitrue\@ifnextchar [{\@tempswatrue
	\@citex}{\@tempswafalse\@citex[]}}
\def\citelow{\@cghifalse\@ifnextchar [{\@tempswatrue
	\@citex}{\@tempswafalse\@citex[]}}
\def\@cite#1#2{{$\null^{#1}$\if@tempswa\typeout
	{IJCGA warning: optional citation argument
	ignored: `#2'} \fi}}

\def\fnt#1#2{\footnotetext{\kern-.3em
	{$^{\mbox{\sevenrm #1}}$}{#2}}}
 1
 1
 1

\font\tenbf=cmbx10
\font\tenrm=cmr10
\font\tenit=cmti10

\font\ninerm=cmr9

\begin{document}

\centerline{\tenbf QUARK--ANTIQUARK BETHE--SALPETER  EQUATION IN QCD }
%\baselineskip=22pt
%\centerline{\tenbf BETHE--SALPETER EQUATION}
\vspace{0.8cm}
\centerline{\tenrm N. BRAMBILLA and G.M. PROSPERI}
\baselineskip=13pt
\centerline{\tenit Dipartimento di Fisica dell'Universit\`a di Milano
 and I.N.F.N.}
\baselineskip=12pt
\centerline{\tenit Via Celoria 16, 20133 Milano}
\vspace{0.9cm}
\abstracts{We review our recent results \cite{bsult}
 on the derivation of a B-S equation in QCD
 in a Wilson loop context. We work  in a second order formalism, use
 the Feynman--Schwinger path integral representation for a quark in
external field and obtain a similar expression for a quark--antiquark
 amplitude in which the gauge fields appears only through the Wilson
 loop integral $W$. A $ q \bar{q}$ B-S equation is obtained starting
 from this expression under the assumption, already used in the
derivation of heavy quark potential, that $i \ln W$ can be expressed as
 the sum of a perturbative  contribution and an area term.
The intrinsically nonperturbative derivation method is first discussed
 on the simpler case of two spinless particles interacting through
 a scalar field.
The $q \bar{q}$ semirelativistic
 potential is reobtained  in the large quark mass limit.}

\vfil
%\vspace{0.8cm}
\rm\baselineskip=14pt
\section{ Introduction}
%\vspace{-0.7cm}
%\subsection{Typeset Scripts}
%\vspace{-0.35cm}

     Many  attempts have been  made
  to apply the Bethe-Salpeter
equation to a study of the spectrum and the properties of the
mesons, hoping
 to obtain an unified and consistent description of the quark-antiquark
bound states envolving light quarks as well as heavy ones. In all
these attempts,
to our knowledge, the choice of
the confinement part of the
 kernel  was purely conjectural
and only made in such a way that the successful heavy quarks potential could be
recovered in the non relativistic limit.
Beside  being theoretically unsatisfactory, as well
known,
 such a
procedure  encounters many conceptual and phenomenological
 difficulties \cite{gara}.\par
On the contrary, in the case of large quark masses it was  possible {\it to
derive} the semirelativistic $q \bar{q}$ potential ``essentially from
first principles'' reducing it to an
 evaluation of the Wilson loop integral \cite{bp,rev}
  \begin{equation}
W = {1\over 3}
 \langle {\rm Tr} {\rm P}_{\Gamma}
 \exp i g \{ \oint_{\Gamma} dx^{\mu} A_{\mu} \}
        \rangle
\label{eq:loop}
  \end{equation}
(the loop $\Gamma$ is supposed
 made by a quark world line ($ \Gamma _1 $),   an
antiquark world line ($ \Gamma _2 $) followed in the
reverse direction and  closed by two
straight lines connecting the initial ($y_1 , y_2$)
 and the final ($x_1, x_2$)  points of the two world
lines; $A_\mu (x)$  denotes a colour
matrix of the form $A_\mu (x) = {1\over 2} A_\mu^a \lambda^a $;
 ${\rm P}_\Gamma $ prescribes the ordering along the loop
 and ${\rm Tr}$  denotes
the trace on the colour
  matrices;
 the expectation value  stands
for a functional  integration
on the gauge field alone  (see below)).

The basic tool for this derivation was the use of a path integral
 representation of the quark propagator in an external field and
the basic  assumption for the evaluation of $W$ was that
the logarithm of this quantity  could be written as the sum of a
perturbative contribution and an area law term
 \begin{equation}
i \ln W = i (\ln W)_{\rm pert} + \sigma S_{\rm min}   .
\label{eq:iniz}
   \end{equation}
In principle
$S_{\rm min}$ should be
 the minimum area enclosed by $\Gamma$;
in practice  one uses
 its straight line approximation, consisting in
replacing it  with the surface spanned by the straight lines connecting
equal time points on the quark and the antiquark worldlines.
Precisely one writes
   \begin{equation}
       S_{\rm min} \cong   \int_{t_{\rm i}}^{t_{\rm f}}
 dt\,  r \int_0^1 d\lambda
 [1-(\lambda {  {\bf v}_{1 {\rm T}} }
           + (1-\lambda) { {\bf v}_{2 {\rm T}}})^2 ]^{1 \over 2},
\label{eq:minform1}
        \end{equation}
($t$ stands for the ordinary time, ${\bf z}
_1 (t)$ and ${\bf z}_2 (t)$ for the quark and the antiquark positions at the
time $t$, ${\bf v}_j = {d {\bf z}_j \over  dt}$ denotes the ordinary
velocity and the suffix ${\rm T}$
specifies  its
  transverse component
  $(\delta^{hk}-{r^h r^k \over
r^2}) {v^k_j}$,
 ${\bf r}(t)$ denotes  the relative position  of the two particles
${\bf z}_1(t)-{\bf
z}_2(t)$). Actually this equation can be shown to be exact up to the  $v^2$
terms.\par
As well known, due to asymptotic freedom, the first term in (2) is
 assumed  to describe correctly the short range  interaction  and
 vanishes as the average distance $\bar{r}$  between the quark and the
antiquark increases;  the second one  can be justified on the basis
of the strong coupling expansion, is assumed to describe the long
range  interaction and vanishes as $\bar{r}$ decreases. Possibly
(2) is  too poor at intermediate  distances and instanton
 contributions  should be taken into account
(attempts already exist in this direction, see e.g. \cite{yn}).
 Furthermore,
 presently the theory is not able  to give a relation  betweeen
 the ``string tension'' $\sigma$
 and the renormalized strong coupling constant
 $\alpha_s = {g^2 \over 4 \pi}$,  which in practice
 have to be treated
as independent variables.  In spite of the above circumstance,
 significant results can be already produced
 by (2) and (3).\par
The difficulty in obtaining a confining Bethe--Salpeter equation
  in QCD comes from  the fact that usually
 this equation is derived by appropriate resummations of
 the perturbative  series expansion
 and obviously such a procedure can provide
 only the perturbative part of the kernel.  In this paper we want
 to review some of our recent results \cite{bsult}
 to show that the technique  used for the potential
 can be  adapted even to the case
 of the   Bethe--Salpeter  equation and  that  a  theoretically
well founded kernel can be obtained in this way on the basis
 of (2) and (3).
Since (3) is not Lorentz invariant,
such equation shall be
  assumed to be valid  in the center of mass  frame.\par
Let us begin to present our results in  more precise terms.
 The QCD lagrangian can be written as
\begin{equation}
L= \sum_{f=1}^{N_f}\bar{\psi}_f (i \gamma^\mu D_\mu -m_f) \psi_f  - {1\over 4 }
 F_{\mu\nu} F^{\mu \nu} + L_{\rm GF}
\label{eq:lagqcd}
\end{equation}
(where $D_{\mu}= \partial_{\mu} -i g A_{\mu}$ and
 $L_{\rm GF}$ is the gauge fixing term),
and as usual the gauge invariant
 quark--antiquark Green function as
\begin{eqnarray}
G^{gi}_4(x_1,x_2,y_1,y_2) &=&
\frac{1}{3}\langle0|{\rm T}\psi_2^c(x_2)U(x_2,x_1)\psi_1(x_1)
\overline{\psi}_1(y_1)U(y_1,y_2)  \overline{\psi}_2^c(y_2)
|0\rangle =
\nonumber\\
&=& \frac{1}{3} {\rm Tr} \langle U(x_2,x_1)
 S^{(1)}(x_1,y_1;A) U(y_1,y_2) C^{-1}
S^{(2)}(y_2,x_2;A) C \rangle .
\label{eq:gauginv}
\end{eqnarray}
Here $c$ denotes the charge-conjugate fields, $C$ is the charge-conjugation
matrix, $U$
the path-ordered gauge string
\begin{equation}
U(b,a)= {\rm P}_{ba}  \exp  \left(ig\int_a^b dx^{\mu} \, A_{\mu}(x) \right)
\label{eq:col}
\end{equation}
(the integration path
 being along  the straight line joining $a$ to $b$),
 $S_1$ and $S_2$ the quark propagators in the
external gauge field $A^{\mu}$
and obviously
\begin{equation}
\langle f[A] \rangle = \frac{\int {\cal D}[A] M_f(A)
f[A] e^{iS[A]}}
{\int {\cal D}[A] M_f(A) e^{iS[A]}} \> ,
\label{eq:med}
\end{equation}
$S[A]$ being the pure gauge field action and $M_f(A)$
the determinant resulting from the explicit integration on the
fermionic fields (which however in practice we set equal to 1 in
 the adopted approximation). \par
The  propagators  $S_1$ and $S_2$ are supposed to be defined
 by  the equation
\begin{equation}
( i\gamma^\mu D_\mu -m) S(x,y;A) =\delta^4(x-y)
\label{eq:propdir}
\end{equation}
and the appropriate  boundary conditions.\par
In the derivation of the potential  one performs over (8)
 a Foldy--Wouthuysen transformation, replaces  the Dirac type
  propagators $S^{(1)}$ and  $S^{(2)}$ by Pauli ones $K^{(1)}$
 and $K^{(2)}$, takes advantage  of ordinary path--integral
 representations for $K^{(1)}$ and $K^{(2)}$. In this way
 one obtains  a two--particle Pauli propagator in which the
 gauge field  appears only through  the quantity $W$ and its
 functional derivatives with respect to the world lines
 of the quark  $q$ and the antiquark $\bar{q}$.
 Finally, after using (2) and (3) and
 expanding in the velocities one  can verify that the obtained propagator
satisfies a two--particle Schr\"odinger equation.\par
For a fully relativistic treatment it  is convenient
 first
to pass to the
second order formalism by setting
\begin{equation}
S(x,y;A) = (i \gamma^\nu D_\nu + m) \Delta_\sigma(x,y;A) ,
\end{equation}
with
\begin{equation}
(D_\mu D^\mu +m^2 -{1\over 2} g \, \sigma^{\mu \nu} F_{\mu \nu})
\Delta_\sigma (x,y;A) = -\delta^4(x-y)
\label{eq:propk}
\end{equation}
($\sigma^{\mu \nu} = {i\over 2} [\gamma^\mu, \gamma^\nu]$). Then,
 taking into account gauge invariance, we can write
\begin{equation}
G_4^{\rm gi}(x_1,x_2; y_1,y_2) =(i \gamma_1^\mu \partial_{x_1 \mu}
 + m_1) ( i \gamma_2^\nu \partial_{x_2 \nu} +m_2) H_4(x_1,x_2;y_1,y_2)
\label{eq:eqg}
\end{equation}
with
\begin{equation}
H_4(x_1,x_2;y_1,y_2) = -{1\over 3} {\rm Tr}
\langle U(x_2,x_1) \Delta_\sigma^{(1)}(x_1,y_1;A)
 U(y_1,y_2) \tilde{\Delta}_\sigma^{(2)}(x_2,y_2;-\tilde{A})\rangle
\label{eq:eqh}
\end{equation}
(where  the tilde denotes  transposition  on the colour indices alone)
and we can use for $\Delta_\sigma^{(1)}$ and
$\Delta_\sigma^{(2)}$ the quadridimensional  path integral
 representation of Feynman--Schwinger
obtaining a corresponding representation
 for $H_4$.\par
Working on such expressions eventually
  we arrive  at
  a "second order"  nonhomogeneous
 Bethe-Salpeter  equation of the form
   \begin{eqnarray}
& &     H_4(x_1,\,x_2;\,y_1,\,y_2) \, = \, H_2^{(1)}(x_1-y_1)\,
H_2^{(2)}(x_2-y_2)
   \,- \,i \int d^4\xi_1 d^4\xi_2 d^4\eta_1 d^4\eta_2 \nonumber \\
 &  &\quad \quad H_2^{(1)}(x_1-\xi_1)\, H_2^{(2)}(x_2-\xi_2)\,
 I(\xi_1,\,\xi_2;\,\eta_1,\,\eta_2) \,
    H_4(\eta_1,\,\eta_2;\,y_1,\,y_2) .
\label{eq:bsh}
\end{eqnarray}
Here $H_2$ stands for a kind of colour independent
 one particle  propagator
 and $I$ denotes the appropriate kernel  obtained
 as an expansion in the strong coupling constant $\alpha_s =
 {g^2 \over 4 \pi}$ and in the string tension $\sigma$ (better
 in $\sigma a^2$, $a$ being  a characteristic length, typically
 the radius of the particular bound state).\par
We stress that the method  by which (13) has been obtained
 is essentially nonperturbative.\par
At the lowest order in $\alpha_s$ and $\sigma a^2$,
 after factorizing the conservation $\delta$,
  we can write in the momentum space
 \begin{equation}
     \hat{I} (p_1,\,p_2;\,p_1^\prime,\,p_2^\prime)\,=\,
\hat I_{\rm pert }
     (p_1,\,p_2;\,p_1^\prime,\,p_2^\prime)\,+\,\hat
 I_{\rm conf} (p_1,
     \,p_2;\,p_1^\prime,\,p_2^\prime)
\label{eq:bform}
 \end{equation}
$(p_1^\prime +p_2^\prime =p_1 +p_2)$,
with
 \begin{eqnarray}
  & & \hat{I}_{\rm pert} =  16 \pi {4 \over 3} \alpha_s
   \{ D_{\rho \sigma}(Q) q_1^\rho
        q_2^\sigma - {i \over 4} \sigma_1^{\mu \nu} (\delta_\mu^\rho Q_\nu-
     \delta_\nu^\rho Q_\mu) q_2^\sigma D_{\rho \sigma } (Q)
+\nonumber \\
 & & \quad \quad +{i \over 4} \sigma_2^{\mu \nu}
        (\delta_\mu^\sigma Q_\nu - \delta_\nu^\sigma Q_\mu) q_1^\rho
  D_{\rho \sigma }(Q)+\nonumber \\
& &
\quad \quad
 +{1 \over 16} \sigma_1^{\mu_1 \nu_1} \sigma_2^{\mu_2 \nu_2}
        (\delta_{\mu_1}^\rho Q_{\nu_1} - \delta_{\nu_1}^\rho Q_{\mu_1})
        (\delta_{\mu_2}^\sigma Q_{\nu_2} - \delta_{\nu_2}^\sigma
Q_{\mu_2})
        D_{\rho \sigma}(Q) \}  +\dots\nonumber \\
& &
\label{eq:ipertq}
\end{eqnarray}

\noindent
   and
  \begin{equation}
     \hat{I}_{\rm conf} = \int d^3 {\bf r} \, e^{i {\bf Q} \cdot {\bf r}}\,
     J({\bf r}, \, q_1, \, q_2) ,
\label{eq:iconf}
\end{equation}
with
 \begin{eqnarray}
   J({\bf r}, \, q_1, \, q_2) & & = {2 \sigma r \over q_{10} + q_{20} }
    \Big [  q_{20}^2 \sqrt{q_{10}^2  -{\bf q}_{\rm T}^2} +
      q_{10}^2 \sqrt  {q_{20}^2 - {\bf q}_{\rm T}^2} + \nonumber \\
     & &+ {q_{10}^2 q_{20}^2 \over \vert {\bf q}_{\rm T} \vert }
      (\arcsin {\vert {\bf q}_{\rm T}\vert \over  q_{10} } +
      \arcsin {\vert {\bf q}_{\rm T}\vert \over  q_{20}  })
\Big ]
+\nonumber \\
& & +2 \sigma  {\sigma^{k \nu}_1 q_{20} q_{1 \nu} r^k \over r \sqrt{q_{10}^2
-
 {\bf q}_{ {\rm T}}^2}} - 2 \sigma
 { \sigma^{k \nu}_2 q_{10} q_{2 \nu} r^k
\over  r \sqrt{q_{20}^2 - {\bf q}^2_{ {\rm T}}} }
+\dots
\label{eq:iconfj}
\end{eqnarray}
In (15)--(17) it has been set
  \begin{equation}
    q_1={p_1 + p_1^\prime \over 2} \, , \ \  \ \ q_2={p_2
      + p_2^\prime \over 2} \,  , \
 \  \  \
      Q = p_1^\prime - p_1   = p_2 - p_2^\prime \,  ,
\label{eq:defq}
\end{equation}
 $D_{\rho \sigma}(Q)$ denotes the usual  gluon free
propagator and
the center of mass frame is understood
 ($ {\bf q}_1 = -{\bf q}_2 = {\bf q} \, , \ q_{\rm T}^h =
(\delta^{hk} - \hat r ^h \hat r ^k) q^k $).\par
   Eqs. (13)-(\ref{eq:iconfj})
 are the main  results of our paper.\par
 Notice, obviously,  that,
instead of (\ref{eq:bsh}) we could have written
the homogeneous equation
 \begin{equation}
     \Phi_P (k) = -i \int {d^4 k^\prime \over (2 \pi)^4} \hat H_2(\eta_1 P
+ k) \hat H_2(\eta_2 P - k) \hat I(k , k^\prime;P) \Phi_P
(k^\prime) \,   ,
\label{eq:fi}
\end{equation}
which is more appropriate for the bound state problem. In this equation $
\eta_j = {m_j \over m_1 + m_2} $, $ P $ denotes
 the total momentum $p_1 + p_2$,
$ k$  stands for the relative momentum $\eta_2 p_1 - \eta_1 p_2 $
($q_j = \eta_j P + { k+k^\prime \over 2}$ and in the CM frame $ {\bf
q} = { {\bf k}^\prime + {\bf k} \over 2}$ ), $\Phi_P(k) $ is
 a kind of Bethe--Salpeter wave function in the momentum space.
\par
{}From (\ref{eq:fi}) by replacing $\hat{H}_2 (p)$  with the free propagator
 $ {-i\over p^2 -m^2}  $ and performing an appropriate instantaneous
 approximation on $\hat{I}$ [consisting in setting:
$ Q_0=0$ and  $q_{j0}= { w_j^\prime + w_j\over 2} $ or $
p_{j0}=p_{j0}^\prime ={ w_j^\prime + w_j\over 2} $ or
 $ k_0=k_0^\prime =\eta_2 { w_1^\prime + w_1 \over 2} -\eta_1
{w_2^\prime + w_2\over 2}$ and $ P_0 ={1\over 2} ( w_1^\prime
+ w_1 + w_2^\prime + w_2 )$, with $w_j=\sqrt{m_j^2 + {\bf k}^2 }$,
 $ w_j^\prime=\sqrt{m_j^2 + {\bf k}^{ \prime 2}}$ ]
  we  can further obtain an
effective square mass operator for the mesons (in the CM frame ${\bf
P}=0, P=(m_B,0) $)
  \begin{equation}
    M^2 = M_0^2 + U
\label{eq:quadr}
  \end{equation}
with  $ M_0 = \sqrt{m_1^2 + {\bf k}^2} + \sqrt{m_2^2 + {\bf k}^2} $ and
\begin{equation}
  \langle {\bf k} \vert U \vert {\bf k}^\prime \rangle =
{1\over (2 \pi)^3 }
 \sqrt{ w_1
  + w_2 \over 2  w_1  w_2}\, \hat I_{\rm inst}(
{\bf k} , {\bf k}^\prime )  \sqrt{ w_1^\prime + w_2^\prime
\over 2  w_1^\prime w_2^\prime}.
\label{eq:quadrrel}
\end{equation}
 The quadratic form of Eq.(\ref{eq:quadr}),
 obviously derives from the second order character of the
used formalism.
  It should be mentioned that for light mesons this form
 seems phenomenologically favoured with respect to the linear one.\par
In more usual terms we can also write
\begin{equation}
        M = M_0 + V
\label{eq:lin}
\end{equation}
with
\begin{equation}
\langle {\bf k} \vert V \vert {\bf k}^\prime \rangle =
{1 \over ( 2 \pi)^3 }
 {1 \over 4 \sqrt{ w_1 w_2 w_1^\prime w_2^\prime } } \hat{I}_{\rm inst}
({\bf k}, {\bf k}^\prime)+\dots
\label{eq:linrel}
\end{equation}
where the dots stand for higher order terms in $\alpha _{\rm s}$ and $\sigma
a^2$  and kinematical factors equal to 1 on the energy shell are
 neglected.
In the limit of small ${{\bf p}^2 \over m^2}$ the potential $V$ as
given by (\ref{eq:linrel})
 reproduces the semirelativistic potential of
 ref. \cite{bp}.
On the contrary, if one neglects
 in $V$ the spin dependent terms, one
 reobtains the hamiltonian of the relativistic flux tube
model \cite{flux} for quarks without spin
 with an appropriate ordering prescription \cite{bp}.\par
     Finally we want to  mention that
 a result
strictly related to our one,  but
 directly
 in hamiltonian form (\ref{eq:lin}),
  has been obtained by
Simonov and collaborators \cite{sim}.\par
The following part of the paper is organized in this way.
In Sect. 2 we illustrate  our nonperturbative method on the model
 case
 of two spinless particles  interacting through a scalar field, to which also
 the usual perturbative derivation applies.
 In Sect.3 we obtain the  mentioned path integral representation
 of $H_4$ and in Sect.4 we sketch a derivation of
 Eqs.(13)-(17). Finally in Sect.5 we discuss how the
 $q \bar{q}$ potential  is reobtained from Eq. (23).

\vskip 1 truecm

\section{Bethe--Salpeter equation for scalar
particles}
Let us consider  two scalar ``material'' fields $\phi_1$ and $\phi_2$
interacting  through a third scalar field $A$ with the coupling
 ${g\over 2} ( \phi_1^2  A -  \phi_2^2 A)$.
Then,  after integration over $\phi_1$ and $\phi_2$,
the full one--particle propagator can be written as
\begin{equation}
G_{2}(x-y)=  \langle 0 \vert {\rm T}\phi (x) \phi (y)\vert
0 \rangle= \langle i\Delta(x, y; A)\rangle \equiv
 {\int {\cal D}A e^{i S_0(A)} M(A) i\Delta (x,y; A) \over
 \int {\cal D}A e^{i S_0(A)} M(A)},
\label{eq:propscal}
\end{equation}
where  $\Delta (x,y,A)$ is the propagator for the
particle
 in the external
 field $A$,
 $S_0(A)$ is the free
 action for the field $A$ and the determinantal factor $M(A)$
 comes from the integration on the fields $\phi$
\begin{eqnarray}
& & M(A)= \prod_{j=1,2}
\Big [{ {\rm det} (\partial^{\mu} \partial_{\mu} +m^2 \mp g A)
\over {\rm det} (\partial^{\mu} \partial_{\mu} +m^2)}\Big ] ^{-{1\over
2}}=1- \nonumber \\
& & -{1\over 2} \sum_{j=1,2} \big \{\mp g \int d^4 x A(x)
\Delta_{\rm F}(0) -{1\over 2} g^2 \int d^4 x d^4 y A(x)
 \Delta_{\rm F}(x-y) A(y) \Delta_{\rm F}(y-x) + \dots \}
\nonumber \\
&& \quad \quad
\label{eq:det}
\end{eqnarray}
(where the upper minus sign refers to the quark , $j=1$ and the lower
 plus sign refers to the antiquark $(j=2)$)
$\Delta_{\rm F}$ denoting the usual scalar particle free propagator.
 \par The covariant
 Feynman-Schwinger representation for $\Delta(x,y;A)$ reads
\begin{eqnarray}
\Delta(x,y; A)& =&
-{ i\over 2} \int_0^{\infty} ds
 \int_{y}^{x} {\cal D} z {\cal D} p\,
{\rm exp}\, \big \{  i\int_0^s {\rm d} \tau [ - p_{\mu }
 \dot{z}^{\mu} +{1\over 2} p_{\mu} p^{\mu }
-{1\over 2} m^2 +{1\over 2} \pm g A(z) ] \big \}
\nonumber\\
& =& -{ i\over 2} \int_0^{\infty} ds
 \int_{y}^{x} {\cal D} z\,
{\rm exp}\,\big \{ - i\int_0^s {\rm d} \tau  {1\over 2} [
 ( \dot{z}^{ 2} +m^2) \mp g  A (z) ] \big \},
\label{eq:propcov}
\end{eqnarray}
where again the upper sign refers to the quark and the lower to the
antiquark; the path integrals are understood to be extended over all
 paths
 $z^{\mu}= z^{\mu}(\tau)$ connecting $y$ with $x$ expressed in
terms of an arbitrary  parameter $ \tau$
with $ 0\le \tau
\le s$. In Eq.(\ref{eq:propcov})
 $\dot{z}$ stands
for ${dz(\tau)
\over d \tau}$ and the ``functional measures'' are assumed to be
defined as
\begin{eqnarray}
 {\cal D} z  & = & ({1\over 2 \pi i \varepsilon })^{ 2 N}
  d^4 z_1 \dots  d^4 z_{N-1}, \quad \qquad \qquad
 {\cal D} p   =  ({i \varepsilon\over 2 \pi })^{ 2 N}
  d^4 p_1 \dots  d^4 p_{N-1} d^4 p_N \nonumber \\
& & {\cal D} z  {\cal D} p  =  ({1\over 2 \pi })^{ 4 N}
 d^4 p_1 d^4 z_1 \dots d^4 p_{N-1} d^4 z_{N-1}
 d^4 p_N.
\label{eq:misfunz}
\end{eqnarray}
($\varepsilon$ being the lattice time spacing and $\varepsilon \to 0$
 being understood).\par
Replacing  Eq.(\ref{eq:propcov}) in (\ref{eq:propscal})
 we obtain
\begin{equation}
G_2 (x-y)= {1\over 2} \int_0^{\infty}ds \int_y^x {\cal D} z
 \exp \{ -{i\over 2} \int_0^s d \tau (\dot{z}^2 +m^2)\}
\langle \exp {\pm i g \over 2} \int_0^{\tau} d \tau A(z)\rangle  ,
\label{eq:propppscal}
\end{equation}
and, if we take  simply $M(A)=1$ (quenched approximation),
\begin{equation}
 \langle \exp { \pm i g \over 2} \int_0^s d \tau
A(z)\rangle   = \exp {i g^2\over 4} \int_0^{s} d \tau
 \int_0^{\tau} d\tau^{\prime}
 D_{\rm F}(z-z^{ \prime}),
\label{eq:av1}
\end{equation}
$D_{\rm F}(x) $ being now the free propagator for field $A$.
\par
 Similarly
for    the two--particle propagator we have
 \begin{eqnarray}
& & G_4(x_1,x_2; y_1, y_2) = \langle 0\vert {\rm T} \phi_1(x_1)
\phi_2(x_2) \phi_1(y_1) \phi_2(y_2) \vert 0\rangle=
\langle G_2^{(1)}(x_1, y_1;A)  G_2^{(2)} (x_2, y_2;A)
\rangle = \nonumber \\
& &
({1\over 2})^2 \int_0^{\infty} ds_1 \int_0^{\infty} ds_2
 \int_{y_1}^{x_1} {\cal D}z_1 \int_{y_2}^{x_2} {\cal D} z_2\times
\exp {-i \over 2} \{ \int_0^{s_1} d \tau_1 (\dot{z}_1^{2}
+m_1^2) +\int_0^{s_2} d \tau_2 (\dot{z}_2^{2}+m_2^2)\}
\nonumber \\
&& \langle
 \exp {i\over 2} \{ g \int_0^{s_1} d \tau_1
 A(z_1) - g \int_0^{s_2} d \tau_2 A(z_2)\}
\rangle
\label{eq:g4}.
\end{eqnarray}
and (always for $M(A)=1 $)
\begin{equation}
 \langle \exp {i\over 2} \{ g \int_0^{s_1} d \tau_1
 A(z_1) - g  \int_0^{s_2} d \tau_2 A(z_2)\}
\rangle =
 \exp \sum_{j=1,2} {i g^2 \over 4} \int_0^{s_j} d \tau_j
 \int_0^{\tau_j} d \tau_j^{\prime} \big [
 D_{\rm F}(z_j - z^{\prime }_j) \big ].
\label{eq:treottobis}
\end{equation}
In conclusion we have
\begin{eqnarray}
& & G_4(x_1,x_2; y_1, y_2)
 =  ({1\over 2})^2 \int_0^{\infty} ds_1 \int_0^{\infty} ds_2
 \int_{y_1}^{x_1} {\cal D}z_1 \int_{y_2}^{x_2} {\cal D} z_2
\nonumber \\
& & \exp {-i \over 2} [ \int_0^{s_1} d \tau_1 (\dot{z}_1^{2}
+m_1^2) -   {g_1^2\over 2} \int_0^{s_1} d \tau_1
 \int_0^{\tau_1} d \tau_1' D_{\rm F}(z_1-z_1')]\times
\nonumber\\
& & \exp {- i \over 2} [
\int_0^{s_2} d \tau_2 (\dot{z}_2^{2}+m_2^2)\}
 -{g_2^2\over 2} \int_0^{s_2} d \tau_2 \int_0^{\tau_2} d \tau_2'
 D_{\rm F}(z_2-z_2') ]\times \nonumber \\
& & \exp  {i g_1 g_2\over 4} \int_0^{s_1} d \tau_1
 \int_0^{s_2} d \tau_2 D_{\rm F}(z_1 -z_2).
\label{eq:g4bis}
\end{eqnarray}
( $z_j $ stands for $z_j(\tau_j)$,
$z_j^\prime $ stands for $z_j(\tau_j^\prime)$).\par
{}From (32),  using
the identity
\begin{eqnarray}
& & \exp {- ig^2 \over 4} \int_0^{s_1} d \tau_1
 \int_0^{s_2} d \tau_2 D_{\rm F}(z_1-z_2)=
 =1 -\nonumber \\
& &- {i g^2 \over 4} \int_0^{s_1} d \tau_1
 \int_0^{s_2} d \tau_2 D_{\rm F}(z_1-z_2) \exp
 [ {-i g^2 \over 4} \int_0^{\tau_1}
 d \tau_1' \int_0^{s_2} d \tau_2' D_{\rm F}(z_1'-z_2')],
\label{eq:id}
\end{eqnarray}
  we can write  after some manipulations
\begin{eqnarray}
& & G_4(x_1, x_2; y_1, y_2) = G_2(x_1-y_1) G_2(x_2-y_2)-
   {i g^2 \over 4}
 ({1 \over 2})^2 \int_0^{\infty} d \tau_1
 \int_{\tau_1}^{\infty} ds_1 \int_0^{\infty} d \tau_2
\int_{\tau_2}^{\infty} ds_2
\nonumber \\
& & \int d^4z_1 \int d^4z_2
\int_{z_1}^{x_1}  {\cal D} z_1
\int_{z_2}^{x_2} {\cal D} z_2 \int_{y_1}^{z_1} {\cal D} z_1
 \int_{y_2}^{z_2} {\cal D} z_2\, D_{\rm F}(z_1-z_2) \nonumber \\
& &  \exp {i\over 2}
 \big \{ \int_{\tau_1}^{s_1} d \tau_1' [ -\dot{z}_1^{'2}-m_1^2
  +{g^2 \over 2} \int_{\tau_1}^{\tau_1'} d \tau_1''
D_{\rm F}(z_1'-z_1'')] +
 \int_{\tau_2}^{s_2} d \tau_2' [ - \dot{z}_2^{'2}-m_2^2
 +{g^2 \over 2} \int_{\tau_2}^{\tau_2'} d \tau_2''\nonumber \\
& & D_{\rm F}(z_2'-z_2'')]\big \}
 \times
\exp{i\over 2} \big \{\int_{0}^{\tau_1} d \tau_1' [ - \dot{z}_1^{'2}-m_1^2
 +{g^2 \over 2} \int_{0}^{\tau_1'} d \tau_1''
D_{\rm F}(z_1'-z_1'')]+ \nonumber \\
& & +\int_{0}^{\tau_2} d \tau_2' [  -\dot{z}_2^{'2}-m_2^2+
\quad  +{g^2 \over 2} \int_{0}^{\tau_2'} d \tau_2''
D_{\rm F}(z_2'-z_2'')]
 -{ g^2 \over 2} \int_0^{\tau_1} d \tau_1'
 \int_0^{\tau_2} d \tau_2' D_{\rm F}
(z_1'-z_2')\big \} \times \nonumber \\
& & \exp i\big \{ {g^2 \over 4}
 \int_{\tau_1}^{s_1} d \tau_1' \int_0^{\tau_1} d \tau_1''
 D_{\rm F}(z_1'-z_1'') +
{g^2 \over 4}
 \int_{\tau_2}^{s_2} d \tau_2' \int_0^{\tau_2} d \tau_2''
 D_{\rm F}(z_2'-z_2'') +\nonumber \\
& & \quad \quad \quad \quad -{g^2 \over 4}
\int_0^{\tau_1} d \tau_1' \int_{\tau_2}^{s_2} d \tau_2'
 D_{\rm F}(z_1'-z_2')\big \}.
\label{eq:orror}
\end{eqnarray}
Then, if we replace
 the last exponential  with 1,
we obtain immediately
the Bethe--Salpeter equation
\begin{eqnarray}
& & G_4(x_1, x_2, y_1, y_2)  =  G_2(x_1-y_1) G_2(x_2-y_2)
  -i \int d^4 z_1 \int d^4 z_2 \int d^4 z_1' \int d^4 z_2'
\nonumber \\
& & G_2(x_1-z_1) G_2(x_2-z_2) I(z_1,z_2; z_1',z_2') G_4(z_1',z_2',y_1,
y_2)
\label{eq:bethesconf}
\end{eqnarray}
with  the {\it ladder approximation} kernel
\begin{equation}
I(z_1,z_2,z_1^\prime,z_2^\prime)= g^2 D_{\rm
F}(z_1-z_2)
\delta^4(z_1-z_1^\prime) \delta^4(z_2-z_2^\prime).
\label{eq:ladder}
\end{equation}
On the contrary,
if we had considered
 the entire
 expansion of the last exponential in (33),
we would have obtained
\begin{eqnarray}
& & I(z_1,z_2,z_1',z_2') = g^2  D_{\rm F}(z_1'-z_2')
\delta^4(z_1-z_1^\prime) \delta^4(z_2-z_2^\prime)
-\nonumber \\
& & -i g^4 \int d^4\xi_1 D_{\rm F}(z_1-z_1')
 G_2(z_1-\xi_1)
 G_2(\xi_1-z_1^\prime)\nonumber \\
& &
 D_{\rm F}(\xi_1-z_2) \delta^4(z_2-z_2')- i
g^4 \int d\xi_2 \delta^4(z_1-z_1')
D_{\rm F}(z_1-\xi_2)
  G_2(z_2-\xi_2)
 G_2(\xi_2-z_2')\nonumber \\
& &   D_{\rm F}(z_2-z_2')
+ i g^4 D_{\rm F}(z_1-z_2^\prime) G_2(z_1-z_1')
D_{\rm F}(z_2-z_1') G_2(z_2-z_2')
+\dots
\label{eq:kerntant}
\end{eqnarray}
(see Fig.1).
Finally one can go beyond  the quenched approximation and
   take into account additional terms in Eq.(25).
 This would
 amount to insert $\phi_1 \bar{\phi}_1$
and $\phi_2 \bar{\phi}_2$  loops
in all
possible  ways inside the graph.\par
Notice that the final form of the kernel  we have obtained is
expressed as an expansion in the coupling constant  $g$
 as in the ordinary derivation. However in the method described,
 once  we have written eq. (31) and set the last exponential in
 (34) equal to 1, Eqs.(35)-(36) follow exactly.
So perturbative expansion appear only at  the level of successive
corrections and not in the basic approximation. This is the reason
 why  the method applies even to QCD when we replace  (31) with
 (2). In fact, as we have already mentioned, we shall see that
 in such a case the kernel
  $I$ is obtained as an expansion both in $\alpha_s$ and $\sigma
a^2$. Obviously by Fourier transform of Eq.(35) we may pass from
this to the  more usual momentum counterpart.\par
\par
For  subsequent  developments  it is important to mention
  that we can also obtain the B-S equation directly in the momentum
space
starting from the
 first line of (26)
 and working
 in the phase space rather than in the configurational one.
  The only difference in such a  case is that we need
   to make explicit reference to the discrete form
  of the path integral (cf.(27)) and use the discrete counterpart
 of (33); we refer to \cite{bsult} for details. In the basic  approximation
 we find
\begin{eqnarray}
& & \tilde{I}(p_1,p_2, p_1^\prime , p_2^\prime) =  4 g^2 \int d^4 z_1
 d^4 z_2 \int {d^4 k_1 \over (2  \pi)^4}
 {d^4 k_2\over (2 \pi)^4} \int d^4
z_1^\prime
 d^4 z_2^\prime e^{-i (p_1 -k_1) z_1} e^{-i
(p_2-k_2)z_2}\nonumber \\
& &    D({z_1+ z_1^\prime\over 2} -
 {z_2+z_2^\prime\over 2}) \,
 e^{i (p_1^\prime-k_1) z_1^\prime} e^{i (p_2^\prime-k_2) z_2^\prime}
\end{eqnarray}
which would literally correspond to
 the mid--point discretization prescription
  (even if immaterial in this case). In (37) $\tilde{I}(p_1,p_2;
p_1^\prime , p_2^\prime )$ is the ordinary Fourier transform
 of $I(z_1,z_2;z_1^\prime, z_2^\prime)$;
introducing the total momentum $P=p_1+p_2$ and the relative
 momentum
 $k= {m_2\over m_1+m_2} p_1 -{m_1 \over m_1+m_2} p_2\equiv
\eta_1 p_1 - \eta_2 p_2$,
we can also factorize the $\delta$ conservation function
\begin{equation}
\tilde{I}(p_1, p_2; p_1^\prime, p_2^\prime) = (2 \pi )^4
\delta^4(P-P^\prime) \hat{I}(k, k^\prime; P)
\label{eq:capdef}
\end{equation}
and write
\begin{equation}
\hat{I}(k,k^\prime;P) = g^2 D_{\rm F}(k-k^\prime).
\label{eq:ladmom}
\end{equation}

\section{ The quark--antiquark propagator}

Let us come back  to the QCD case.\par
 The Feynman--Schwinger
representation can now be written
\begin{eqnarray}
& & \Delta^\sigma (x,y;A)  =
- {i\over 2}\int_0^\infty ds  {\rm P}_{xy} {\rm T}_{xy}
 \exp {i s\over 2}
 ( - D_\mu D^\mu - m^2 + {1\over 2} g \sigma^{\mu \nu} F_{\mu \nu})
=
 -{i\over 2} \int_0^\infty d s
\nonumber \\
&& \int_y^x {\cal D}z\,
  {\rm P}_{xy} {\rm T}_{xy} {\rm exp}\, i \int_0^{s}
 d\tau \{-{1\over 2} (m^2 +\dot{z}^2) + g A_\rho (z) \dot{z}^\rho
 + {g\over 4} \sigma^{\mu \nu} F_{\mu \nu}(z) \} ,
\label{eq:part}
\end{eqnarray}
where   ${ \rm P}_{xy}$ and ${\rm T}_{xy}$  prescribe the ordering
along the path
 of the colour  and of the spin matrices respectively.
\par
Furthermore, notice that,
 as a consequence of
 a variation in the path $z^\mu(\tau)
\to z^\mu(\tau)+ \delta z^\mu(\tau)$
respecting the extreme points, one has
\begin{eqnarray}
\delta & &\big \{ {\rm P}_{xy} \exp ig \int_0^s d\tau \dot{z}^\mu(\tau) A_{\mu}
(z)\big \}=\nonumber \\
& & = ig \int_0^s \delta S^{\mu\nu}(z(\tau)) {\rm P}_{xy}
\big \{- F_{\mu \nu}(z(\tau)) \exp ig \int_0^s d\tau^\prime
\dot{z}^\mu(\tau^\prime) A_{\mu}(z(\tau^\prime)) \big \}
\label{eq:varwils}
\end{eqnarray}
with $\delta S^{\mu \nu}(z)= {1\over 2} (d z^\mu \delta z^\nu- dz^\nu
\delta z^\mu)$.
 So one  can    write
\begin{eqnarray}
{\rm T}_{xy}& &  \exp(-{1\over 4} \int_0^s d\tau \sigma^{\mu\nu} {\delta
\over \delta S^{\mu \nu}(z) } )
\Big ( {\rm P}_{xy} \exp ig \int_0^s d \tau^\prime \dot{z}^\mu(\tau^\prime)
A_{\mu}(z(\tau^\prime)) \Big )\nonumber \\
& & = {\rm T}_{xy} {\rm P}_{xy} \exp ig \int_0^s d\tau [
\dot{z}^\mu(\tau)A_{\mu}(z(\tau)) + {1\over 4} \sigma^{\mu \nu}
F_{\mu \nu}(z(\tau )) ]
\label{eq:deffmu}
\end{eqnarray}
 and
Eq.(\ref{eq:part}) can  be rewritten as
\begin{equation}
\Delta^\sigma(x,y; A)= -{i\over 2} \int_0^s d\tau \int_y^x {\cal
D}z  {\rm P}_{x y}
{\rm T}_{xy} {\cal S}_0^s \exp \, i \int_0^s d\tau [-{1\over 2} (m^2
+\dot{z}^2)
+ ig\dot{\bar{z}}^\mu A_{\mu}(\bar{z})]
\label{eq:partbis}
\end{equation}
with
\begin{equation}
{\cal S}_0^s =  \exp \Big [ -{1\over 4} \int_0^s d \tau
 \sigma^{\mu \nu} {\delta \over \delta S^{\mu \nu}(\bar{z})}
\Big ].
\label{eq:defop}
\end{equation}
In (44)
 it is understood that $ \bar{z}^\mu(\tau)$ has to be put
equal to  $z^\mu(\tau)$ after the action of ${\cal S}_0^{s} $.
 Alternatively, it is convenient
to write $\bar{z}= z+\zeta$, assume that
${\cal S}_0^s$ acts on $\zeta (\tau)$ with
$ \delta S^{\mu \nu} (z) = {1\over 2} ( d z^\mu \delta \zeta^\nu
 - d z^\nu \delta \zeta^\mu )$ and set eventually $\zeta =0$.\par
 Replacing (44) in (12) we obtain
\begin{eqnarray}
H_4(x_1,x_2;y_1,y_2) & & = ({1 \over 2})^2 \int_0^{\infty} d s_1
\int_0^{\infty} d s_2
 \int_{y_1}^{x_1} {\cal D} z_1\int_{y_2}^{x_2} {\cal D} z_2
 {\rm T}_{x_1 y_1} {\rm T}_{x_2 y_2}
 {\cal S}_0^{s_1} {\cal S}_0^{s_2} \nonumber \\
& & \exp{ ({-i \over 2}) \big \{
 \int_0^{s_1} d\tau_1 (m_1^2 +\dot{z}_1^2) + \int_0^{s_2}
d\tau_2 (m_2^2 +\dot{z}_2^2)\big \} }\nonumber \\
& & {1\over 3} \langle {\rm Tr} {\rm P}_\Gamma
 \exp (ig) \big \{ \oint_{\Gamma} d\bar{z}^\mu A_{\mu}
(\bar{z})   \}
\rangle    ,
\label{eq:hquatr}
\end{eqnarray}
where now $ \bar{z}= \bar{z}_j= z_j +\zeta_j $ on $\Gamma_1 $ and
$\Gamma_2$, $ \bar{z}=z $ on the end lines $x_1 x_2$ and $y_2 y_1$
 and  again the final limit $\zeta_j \to 0$ is   understood.\par
Then, let us try to be more explicit concerning Eqs. (2) and (3).
For the first term in (2) we have,
 at the lowest order of perturbation theory,
\begin{eqnarray}
& &i (\ln W)_{\rm pert} = i
 \ln \langle {1\over 3} {\rm Tr} {\rm P} \exp ig \oint_{\Gamma} dz^{\mu}
 A_{\mu}(z)\rangle_{\rm pert} ={4\over 3} g^2 \int_0^{s_1} d \tau_1
\int_0^{s_2} d\tau_2 D_{\mu \nu}(z_1-z_2)
 \dot{z}_1^{\mu}
\dot{z}_2^{\nu}- \nonumber \\
& & -{2\over 3} g^2 \int_0^{s_1} d \tau_1
\int_0^{s_1} d\tau_1^{\prime}
 D_{\mu \nu}(z_1-z_1^{\prime}) \dot{z}_1^{\mu}
\dot{z}_1^{\prime\nu}-
{2\over 3} g^2 \int_0^{s_2} d \tau_2
\int_0^{s_2} d\tau_2^{\prime} D_{\mu
\nu}(z_2-z_2^{\prime}) \dot{z}_2^{\mu}
\dot{z}_2^{\prime \nu}+ \dots
\label{eq:wilpert}
\end{eqnarray}
On the other side, for the second term in general we have to write
\begin{equation}
S_{\rm min} = \int_{t_i}^{t_f} dt
 \int_0^1 d\lambda
 \big [ -({\partial u^\mu \over \partial t} {\partial u_\mu
\over \partial t })({\partial u^\mu \over \partial \lambda} {\partial u_\mu
\over \partial \lambda })+
 ({\partial u^\mu \over \partial t} {\partial u_\mu
\over \partial \lambda})^2 \big ]^{1\over 2}
\label{eq:smin}
\end{equation}
$x^\mu = u^\mu (t,\lambda) $ being  the equation of the minimal surface
with  contour $\Gamma$. Let us  assume  that for fixed $t$ and for
$\lambda$
varying from 0 to 1, $u^\mu (\lambda,t)$ describes a line connecting a point
on the quark world line $\Gamma_1$  with one on the antiquark
 world line $\Gamma_2$,
\begin{equation}
u^\mu(1,t)= z_1^\mu(\tau_1(t)),\quad \quad \quad
u^\mu(0,t)= z_2^\mu(\tau_2(t)).
\label{eq:udef}
\end{equation}
Obviously (\ref{eq:smin}) is invariant under reparametrization, so
 a priori the parameter  $t$  could be everything.
In particular, however, if  $\Gamma_1$ and $\Gamma_2$
never go backwards in time,
  $t$  can be choosen
as the ordinary time, as in (3),  $u^0(s,t)\equiv t$.
Then
$\tau_1 (t)$ and $\tau_2(t)$ are specified by the equation
\begin{equation}
z_1^0(\tau_1) =z_2^0 (\tau_2)= t .
\label{eq:zzero}
\end{equation}
We shall set
\begin{equation}
L=\int_0^1 d\lambda
\big [ -({\partial u^\mu \over \partial t} {\partial u_\mu
\over \partial t })({\partial u^\mu \over \partial \lambda} {\partial u_\mu
\over \partial \lambda })+
 ({\partial u^\mu \over \partial t} {\partial u_\mu
\over \partial \lambda})^2 \big ]^{1\over 2}.
\label{eq:defl}
\end{equation}
Obviously  $L$  cannot depend only on
 on the extremal points
 $z_1(\tau_1)$ and $z_2(\tau_2)$, but has to depend
 even on the shape of the
 world lines, at least in a neighbourhood of such points. So,
we can think of it as  a function  of $z_1$, $z_2$ and
of all their derivatives in $\tau_1$
 and $\tau_2$,
$L=L(z_1,z_2, \dot{z}_1,
\dot{z}_2, \dots )$.
Finally we can  write (48) as (recall $\dot{z}_j= {d z_j\over
d\tau_j}$)
\begin{equation}
S_{\rm min}= \int dz_1^0 \int dz_2^0 \delta (z_1^0-z_2^0)
L(z_1,z_2,\dot{z}_1,\dot{z}_2, \dots)=
\int d\tau_1 \int d\tau_2 \delta(z_1^0-z_2^0) \dot{z}_1^0 \dot{z}_2^0
L(z_1,z_2,\dot{z}_1,\dot{z}_2, \dots ).
\label{eq:deflmin}
\end{equation}
 In principle this expression can be
considered a good approximation
  even if
 the world lines contain pieces going backwards in time. In fact,
in such a case
  if we fix e.g. $\tau_1$,
 (\ref{eq:zzero})  has more than one solution in
$\tau_2$ and, if $\Gamma_1$ and $\Gamma_2$ are not too much irregular
in space (otherwise $S_{\rm min}$ is large and the weight of the loop
 is small),
the minimal surface can be reconstructed as the algebraic
 sum of various pieces of surface.\par
In  the straight line approximation we must choose
\begin{equation}
u^0(\lambda,t)  =  t \quad ;\quad
u^k(\lambda,t)  = \lambda z_1^k (\tau_1(t)) + (1-\lambda) z_1^k(\tau_2(t))
\label{eq:strai}
 \end{equation}
and
 we have
\begin{equation}
 \dot{z}_1^0 \dot{z}_2^0 L=  \sigma
  \vert {\bf z}_1 -{\bf z}_2\vert
\int_0^1 d\lambda
 \big \{ {\dot{z}_{10}^{ 2}} {\dot{z}_{20}^{ 2}} -
 (\lambda \dot{\bf z}_{1{\rm T}} \dot{z}_{20}
 + (1-\lambda) \dot{\bf z}_{2 {\rm T}}
 \dot{z}_{10} )^2 \big \}^{1\over 2}
\label{eq:lstrai}
\end{equation}
which, introduced in (52) becomes  equivalent to Eq.(3).
The important point  is that (52) with (54)  has the
same general  form  as (47).
However we stress that the  approximation (53) must
  be  performed only {\it
after} that the application of the operators ${\cal S}_0^{s_1}$
and ${\cal S}_0^{s_2}$ has been performed.\par
Substituting (\ref{eq:wilpert}) and (52) in
(\ref{eq:hquatr})
 we obtain at the lowest order
\begin{eqnarray}
& & H_4(x_1,x_2;y_1,y_2)  =  ({1 \over 2})^2 \int_0^{\infty} ds_1
\int_0^{\infty} ds_2
 \int_{y_1}^{x_1} {\cal D} z_1\int_{y_2}^{x_2} {\cal D} z_2
{\rm T}_{x_1 y_1} {\rm T}_{x_2 y_2}
 {\cal S}_0^{s_1} {\cal S}_0^{s_2} \nonumber \\
& & \exp i  \big \{ -{1\over 2}
 \int_0^{s_1} d\tau_1 (m_1^2 +\dot{z}_1^2) -{1\over 2} \int_0^{s_2}
d\tau_2 (m_2^2 +\dot{z}_2^2)+ \nonumber \\
& &  +{2\over 3} g^2 \int_0^{s_1} d\tau_1
\int_0^{s_2} d\tau_1^\prime D_{\mu \nu}(\bar{z}_1-\bar{z}_1^\prime)
\dot{\bar{z}}_1^\mu \dot{\bar{z}}_1^{\nu\prime}+
{2\over 3} g^2
\int_0^{s_2} d\tau_2 \int_0^{s_2} d\tau_2^\prime D_{\mu \nu}(\bar{z}_2
-\bar{z}_2^\prime) \dot{\bar{z}}_2^\mu \dot{\bar{z}}_2^{\nu\prime}
\nonumber \\
& &
- \int_0^{s_1} d\tau_1 \int_0^{s_2} d\tau_2
E(\bar{z}_1, \bar{z}_2, \dot{\bar{z}}_1, \dot{\bar{z}}_2,\dots)\big
\},
\label{eq:hpath}
\end{eqnarray}
where we have set
\begin{eqnarray}
E(z_1, z_2, \dot{z}_1, \dot{z}_2 \dots ) = & & {4\over 3} g^2
 D_{\mu \nu} (z_1-z_2) \dot{z}_1^\mu \dot{z}_2^\nu + \nonumber \\
 & & + \sigma \delta(z_{10} - z_{20}) \dot{z}_{10} \dot{z}_{20}
 L(z_1, z_2, \dot{z}_1, \dot{z}_2 \dots ).
\label{eq:defe}
\end{eqnarray}
\indent Now, let
 us  denote   the quantity in
 curly bracket in (55)  by $\Phi$
 and
  perform a Legendre transformation by introducing the momenta
  $ p_{j \mu}=-
{\delta \Phi\over \delta \dot{z}_{j}^{\mu}}$
( where the various quantities $z_j$, $\dot{z}_j$,
 $\ddot{z}_j, \dots $ are assumed to be treated as
 independent)
\begin{eqnarray}
 { p}_{\mu 1}
 & =& \dot{{ z}}_{\mu 1} + {4\over 3} g^2\int_0^{s_1}
 d\tau_1^\prime D_{\mu \nu}(\bar{z}_1-\bar{z}_1^{\prime})
\dot{\bar{z}}_1^{\prime\nu} + \int_0^{s_2} d\tau_2^\prime {
\partial E( \bar{z}_1, \bar{z}_2^\prime. \dot{\bar{z}}_1 ,
\dot{\bar{z}}_2 \dots ) \over \partial \dot{z}_1^\mu}
\nonumber\\
 {p}_{\mu 2} & =& \dot{z}_{\mu 2} +{4\over 3} g^2 \int_0^{s_2}
 d\tau_2^\prime D_{\mu \nu}(z_2-z_2^{\prime})
\dot{\bar{z}}_2^{\prime\nu}+
 \int_0^{s_1} d\tau_1^\prime {
\partial E( \bar{z}_1, \bar{z}_2^\prime. \dot{\bar{z}}_1 ,
\dot{\bar{z}}_2 \dots ) \over \partial \dot{z}_2^\mu}.
\label{eq:momqq}
\end{eqnarray}
Eq.(\ref{eq:momqq}) cannot be inverted in a closed form
 with respect to $\dot{z}_1$ and $\dot{z}_2$,
 however, we can
do this by an expansion in $\alpha_s= {g^2 \over 4 \pi }$ and
${\sigma a^2}$ and at the lowest order we have
\begin{eqnarray}
\dot{z}_1^{\mu}& =&  p_1^{ \mu} -
{4\over 3} g^2 \int_0^{s_1} d \tau_1^\prime  D_{\mu
\nu}(\bar{z}_1 -z_1^{\prime}) \bar{p}_1^{\prime \nu}
- \int_0^{s_2} d\tau_2^\prime {
\partial E( \bar{z}_1, \bar{z}_2^\prime, \bar{p}_1 ,
\bar{p}_2^\prime  \dots ) \over \partial p_1^\mu}
+\dots
\nonumber \\
\dot{z}_2^{\mu}& =&  p_2^{\mu}
-{4\over 3} g^2 \int_0^{s_2} d \tau_2^\prime D_{\mu
\nu}(\bar{z}_2- \bar{z}_2^{\prime}) p_2^{\prime \nu}-
 \int_0^{s_1} d\tau_1^\prime {
\partial E( \bar{z}_1, \bar{z}_2^\prime, \bar{p}_1^\prime ,
\bar{p}_2  \dots ) \over \partial p_2^\mu}
+\dots
\label{eq:defv}
\end{eqnarray}
with
\begin{equation}
\bar{p}_j^\mu = p_j^\mu + \dot{\zeta}_j^\mu .
\label{eq:defpvar}
\end{equation}
In conclusion we find (up to a determinantal factor that in
 this approximation can be set equal to 1)
\begin{eqnarray}
 & & \quad \quad H_4(x_1,x_2,y_1,y_2) = ({1 \over 2})^2 \int_0^\infty ds_1
\int_0^\infty ds_2 \int_{y_1}^{x_1} {\cal D}z_1 {\cal D}p_1
 \int_{y_2}^{x_2} {\cal D}z_2 {\cal D}p_2 T_{x_1 y_1} T_{x_2 y_2}
\nonumber \\
& &  {\cal S}_0^{s_1} {\cal S}_0^{s_2}
  \exp i \Big \{ \int_0^{s_1} d\tau_1 K_1 + \int_0^{s_2} d\tau_2
 K_2   -  \int_0^{s_1} d\tau_1 \int_0^{s_2} d\tau_2
E(\bar{z}_1, \bar{z}_2, \bar{p}_1, \bar{p}_2, \dots )+\dots \Big \}
\nonumber \\
& & \quad \quad ,
\label{eq:defhe}
\end{eqnarray}
where
\begin{equation}
K_j= -p_j\cdot \dot{z}_j +{1\over 2} (p_j^2 -m_j^2)
+{2\over 3} g^2 \int_0^{s_j} d\tau_j^\prime D_{\mu \nu} (\bar{z}_j
-\bar{z}_j^\prime) \bar{p}_j^\mu \bar{p}_j^{\nu\prime}+\dots
\label{eq:defk}
\end{equation}
includes the self--interacting term.  Notice that now in ${\cal
S}_0^{s_j}$ it  must be understood  $ \delta S^{\mu \nu}
(\bar{z}_j) ={1\over 2} d \tau_j (p_j^\mu \delta \zeta_j^\nu -p_j^\nu
\delta \zeta_j^\mu ) +\dots$. \par

\section{The Bethe--Salpeter equation in QCD}

{}From  Eq. (\ref{eq:defhe}) we proceed along the same line followed in
Sect. 2. with reference to Eq.(32). \par
Using
\begin{eqnarray}
&& \quad \quad \exp  \int_0^{s_1} d\tau_1 \int_0^{s_2} d \tau_2
 E(\bar{z}_1, \bar{z}_2, \bar{p_1}, \bar{p_2} \dots ) =\nonumber \\
=& & 1 +  \int_0^{s_1} d \tau_1 E(\bar{z}_1, \bar{z}_2, \bar{p}_1 ,
\bar{p}_2 \dots) \exp \int_0^{\tau} d\tau^\prime E(\bar{z}_1^\prime,
\bar{z}_2^\prime, \bar{p}_1^\prime, \bar{p}_2^\prime,\dots )
\label{eq:id}
\end{eqnarray}
corresponding to (33),
 we have
\begin{eqnarray}
& & H_4(x_1,x_2;y_1,y_2) = ({1 \over 2})^2
\int_0^\infty ds_1
\int_0^\infty ds_2 \int_{y_1}^{x_1} {\cal D}z_1 {\cal D}p_1
 \int_{y_2}^{x_2} {\cal D}z_2 {\cal D}p_2
\nonumber \\
& & T_{x_1 y_1} T_{x_2 y_2}  {\cal S}_0^{s_1} {\cal S}_0^{s_2}
 \bigg \{ \exp i [ \int_0^{s_1} d\tau_1 K_1 + \int_0^{s_2} d\tau_2
 K_2 ]- i \int_0^{s_1} d \tau_1 \int_0^{s_2} d\tau_2
 E(\bar{z}_1, \bar{z}_2, \bar{p}_1, \bar{p}_2\dots )
\nonumber \\
& &\times  \exp i \Big \{  \int_0^{s_1} d\tau_1 K_1 + \int_0^{s_2} d\tau_2
 K_2
 - \int_0^{\tau_1} d\tau_1^\prime
\int_0^{s_2} d\tau_2^\prime
 E(\bar{z}_1^\prime, \bar{z}_2^\prime,
\bar{p}_1^\prime, \bar{p}_2^\prime, \dots)  \bigg \}.
\label{eq:hlav}
\end{eqnarray}
To obtain  from this an equation analogous to (34) we need to
commute
 ${\cal S}_0^{s_1} {\cal S}_0^{s_2}$ with $E$.
To this aim using the method of Ref. \cite{bsult} and bearing in mind
 (\ref{eq:defe}), (\ref{eq:defl}) and (\ref{eq:hpath}),
  we find first
\begin{eqnarray}
 {\delta \over \delta S^{\mu \nu}(z_1)} & &  \int_0^{s_1} d
\tau_1^\prime \int_0^{s_2} d \tau_2^\prime E(z_1^\prime, z_2^\prime,
p_1^\prime, p_2^\prime, \dots) =\nonumber \\
& & = \int_0^{s_2} d\tau_2^\prime \Big [ {4\over 3} g^2 \big (
\partial_\nu D_{\mu \sigma} (z_1-z_2^\prime) -\partial_{\mu} D_{\nu
\sigma}(z_1-z_2^\prime) \big )p_2^\sigma +\nonumber \\
& & + \sigma \delta(z_{10}-z_{20}) { p_{1\nu} (z_{1\mu}
-z_{2\mu}^\prime ) - p_{1\mu} (z_{1\nu} -z_{2\nu}^\prime )
\over \sqrt{ ({p}_{10}^2 -\dot{\bf p}_1^2 ) ( {\bf z}_1- {\bf z}_2 )^2
+ ( {\bf p}_1 \cdot ({\bf z}_1- {\bf z}_2^\prime))^2 }}
 +\dots  \Big ]
\label{eq:varnonpert}
\end{eqnarray}
and a similar result, with a minus sign  in front, for the
derivative  $ {\delta \over \delta S^{\mu \nu} (z_2^\prime)}$.
Furthermore
\begin{equation}
{\delta^2 \over   \delta S^{\mu \nu} (z_1)  \delta S^{\rho \sigma}
(z_1^\prime) } \int_0^{s_1} d\tau_1^{\prime \prime} \int_0^{s_2}
 d\tau_2^{\prime \prime} E = {\delta^2 \over \delta S^{\mu \nu}(z_2)
\delta S^{\rho \sigma}(z_2^\prime ) }
\int_0^{s_1} d\tau_1^{\prime \prime} \int_0^{s_2} d\tau_2^{\prime
\prime} E= 0,
\label{eq:varzero}
 \end{equation}
but
\begin{equation}
{\delta^2 \over \delta S^{\mu_1 \nu_1} (z_1) \delta S^{\mu_2\nu_2}
(z_2) } \int_0^{s_1} d\tau_1^{\prime \prime} \int_0^{s_2}
d\tau_2^{\prime \prime} E = {4\over 3} g^2
(\delta_{\mu_1}^\rho \partial_{\nu_1} -\delta_{\nu_1}^\rho
\partial_{\mu_1} ) ( \delta_{\mu_2}^\sigma \partial_{\nu_2}
-\delta_{\nu_2}^\sigma \partial_{\mu_2} ) D_{\rho \sigma} (z_1-z_2).
\label{eq:varpert}
\end{equation}
\indent Then,
taking into account the relation
\begin{equation}
e^A B e^{-A} =\sum_{n=0}^\infty {1\over n !} [A,[A, \dots [A,B] \dots
]],
\label{eq:idcom}
\end{equation}
 we have
\begin{eqnarray}
 & &
 \int_0^{s_1} d\tau_1 \int_0^{s_2} d\tau_2
 {\cal S}_{\tau_1-\varepsilon}^{\tau_1 +\varepsilon}
 {\cal S}_{\tau_2-\varepsilon}^{\tau_2+\varepsilon}
    E(\bar{z}_1, \bar{z}_2, \bar{p}_1,
\bar{p}_2, \dots )  ({\cal S}_{\tau_1-\varepsilon}^{\tau_1+\varepsilon}
 {\cal S}_{\tau_2-\varepsilon}^{\tau_2+\varepsilon} )^{-1}= \nonumber \\
& &= (1-{1\over 4} \int_0^{s_1} d s_1^\prime \sigma_1^{\mu_1 \nu_1}
{\delta \over \delta S^{\mu_1 \nu_1}(\bar{z}_1^\prime )})
(1-{1\over 4} \int_0^{s_2} d s_2^\prime \sigma_2^{\mu_2 \nu_2}
{\delta \over \delta S^{\mu_2 \nu_2}(\bar{z}_2^\prime )})
\nonumber \\
& & \int d\tau_1 \int d\tau_2
  E(\bar{z}_1, \bar{z}_2, \bar{p}_1,
\bar{p}_2, \dots )= R(z_1, z_2, p_1,p_2)
\label{eq:defr}
\end{eqnarray}
with
\begin{equation}
R  = R_{\rm pert} + R_{\rm conf}
\end{equation}
\begin{eqnarray}
& & \quad  R_{\rm pert}=   -{4\over 3} g^2  \Big \{
 D_{\rho \sigma} (z_1-z_2) p_1^\rho p_2^\sigma \nonumber \\
& & -{1\over 4} \sigma_1^{\mu \nu} (\delta_{\mu}^{\rho} \partial_{1\nu}
-\delta_\nu^\rho \partial_{1\mu} ) D_{\rho \sigma}(z_1-z_2)p_2^\sigma
- {1\over 4}\sigma_2^{\mu \nu} (\delta_{\mu}^{\sigma} \partial_{2\nu}
-\delta_\nu^\sigma \partial_{2\mu} ) D_{\rho \sigma}(z_1-z_2)p_1^\rho
\nonumber \\
& & +{1\over 16} \sigma_1^{\mu_1 \nu_1} \sigma_2^{\mu_2 \nu_2}
 (\delta_{\mu_1}^\rho \partial_{1\nu_1} -\partial_{\nu_1}^\rho
\partial_{1\mu_1} )
(\delta_{\mu_2^\sigma}\partial_{2\nu_2} -\partial_{\nu_2}^\sigma
\partial_{2\mu_2} )  D_{\rho \sigma}(z_1- z_2)\Big \}
\label{eq:rpert}
\end{eqnarray}
and
\begin{eqnarray}
& & R_{\rm conf}  =  \sigma \delta(z_{10}-z_{20}) \bigg \{
\vert {\bf z}_1- {\bf z}_2 \vert \int_0^1  ds \sqrt{p_{10}^2 p_{20}^2
-[s p_{1{\rm T}} p_{20} + (1-s) p_{2 {\rm T}} p_{10} ]^2 }
\nonumber \\
&& -{1\over 4} p_{20} \sigma_1^{\mu\nu} { p_{1\nu} (z_{1\mu}
-z_{2\mu})  -p_{1\mu} ( z_{1\nu} -z_{2\nu}) \over
 \vert {\bf z}_1 -{\bf z}_2 \vert \sqrt{p_{10}^2 -{\bf p}_{1 {\rm
T}}^2 }}
 +{1\over 4} p_{10} \sigma_2^{\mu\nu} { p_{2\nu} (z_{1\mu}
-z_{2\mu}) -p_{2\mu} ( z_{1\nu} -z_{2\nu}) \over
 \vert {\bf z}_1 -{\bf z}_2 \vert \sqrt{p_{20}^2 -{\bf p}_{2 {\rm
T}}^2 }}\bigg \}\nonumber \\
& & \quad \quad .
\label{eq:rspez}
\end{eqnarray}
Notice that in (68) we have eventually supressed reference to the
higher order derivatives in $z_1, z_2$ and this  corresponds to the
adoption of the straight line approximation.\par
Finally setting
\begin{equation}
H_2(x-y) ={-i\over 2}\int_0^\infty ds
 \int_y^x {\cal D}z {\cal D}p\, T_{xy} {\cal S}_0^s \exp i
\int_0^s d\tau K
\label{eq:hdue}
\end{equation}
we can write Eq.(18) as
\begin{eqnarray}
& & \quad \quad H_4(x_1,x_2; y_1,y_2)  = H_2 (x_1-y_1) H_2(x_2-y_2)+
\nonumber\\
& & - {i\over 4} \int_0^{\infty} d s_1 \int_0^\infty d s_2
 \int_{y_1}^{x_1} {\cal D}z_1 {\cal D}p_1 \int_{y_2}^{x_2}
 {\cal D} z_2 {\cal D} p_2  {\rm T}_{x_1 y_1} {\rm T}_{x_2 y_2}
 \int_0^{s_1}d\tau_1
 \int_0^{s_2} d\tau_2  R(z_1, z_2, p_1,p_2) \nonumber \\
& & {\cal S}_0^{s_1} {\cal S}_0^{s_2}  \exp i \Big \{ \int_0^{s_1}
 d\tau_1^\prime  K_1^\prime + \int_0^{s_2} d\tau_2^\prime
 K_2^\prime -i \int_0^{\tau_1} d\tau_1^\prime \int_0^{s_2}
 d\tau_2^\prime  E(z_1^\prime, z_2^\prime, p_1, p_2 \dots )\Big \}.
\label{eq:hquatrr}
\end{eqnarray}
\par At this point,
as mentioned at the end of Sect.2, to go ahead
 it is necessary to take explicitly into account the
discrete  form  of (\ref{eq:hquatrr}). If, in particular,
 we set
\begin{equation}
{\rm P} \exp  \big [i g \oint_{\Gamma} dz^\mu A_{\mu}(z) \big ]
 = {\rm P}
\prod_{\Gamma}  U(z_n, z_{n-1}) = {\rm P} \exp i g
 \sum_{\Gamma} (z_n^\mu - z_{n-1}^\mu ) A_\mu ( {z_n + z_{n-1}\over 2}
)
\label{eq:wilsdiscr}
\end{equation}
(as required by a gauge invariant  definition of the integral on the
gluon field)  by  appropriate manipulations we eventually obtain
 (13) with (cf.(38))
\begin{eqnarray}
& & \quad \quad \quad (2 \pi )^4 \delta (p_1 +p_2 - p_1^\prime
-p_2^\prime )
\hat{I} ( p_1, p_2 ; p_1^\prime, p_2^\prime) =
  - 4 i\int d^4 \xi_1  d^4 \xi_2 \int d^4 \eta_1 d^4 \eta_2
\nonumber \\
& & \int { d^4 k_1 \over ( 2 \pi )^4 } {d^4 k_2 \over ( 2 \pi )^4 }
e^{ i (p_1 - k_1 ) \xi_1 + i ( p_2 - k_2 ) \xi_2 }
  R( { \xi_1 +\eta_1 \over 2}, { \xi_2 +\eta_2 \over 2}, k_1, k_2 )
 e^{ - i (p_1^\prime -k_1 )\eta_1 -i (p_2^\prime -k_2) \eta_2 }
\nonumber \\
&& \quad \quad .
\label{eq:kernp}
\end{eqnarray}
Then, using
 (69)--(71) and performing explicitely
 the integrals
we find Eqs. (14)--(18).\par
Obviously the homogeneous Eq. (19)
follow from (13)
using the usual decomposition
of the  quark--antiquark propagator in terms of Bethe-Salpeter
 wave functions
\begin{equation}
G_4(k, k^\prime, P) = \sum_B{ \Psi_B(k) \bar{\Psi} (k^\prime)
\over P^2 -m_B^2 }+ {\rm regular}\,  {\rm  terms}.
\end{equation}

\section{ The semirelativistic potential}
{}From (23), expanding in ${1\over m}$  and passing to the coordinate
 representation, we obtain the following  potential
\begin{eqnarray}
V & & = {4\over 3} {\alpha_s \over r} + \sigma r
+ {4 \over 3} {\alpha_s \over m_1 m_2} \big \{ {1 \over 2 r}
 (\delta^{hk} + {r^h r^k \over r^2 }) q^h q^k \big \}_{\rm
W}- \nonumber \\
& & - {4\over 3} i \alpha_s ( {1\over 2 m_1 } {\alpha_1 \cdot {\bf r}
 \over r^3 } - {1\over 2 m_2 } { \alpha_2 \cdot {\bf r} \over r^3})
+ {4\over 3} {\alpha_s \over 2 m_1 m_2} ({\bf \sigma}_1 + {\bf
\sigma}_2 ) \cdot ( {\bf r} \times {\bf q} ) +\nonumber \\
& & + {1\over 3} {\alpha_s \over m_1 m_2 } [ { 3 ( \sigma_1 \cdot {\bf
r} ) (\sigma_2 \cdot {\bf r}) \over r^5 } - {\sigma_1 \cdot \sigma_2
\over r^3  } ] + {4\over 3} {\alpha_s \over m_1 m_2 }
 { 2 \pi \over 3} (\sigma_1 \cdot \sigma_2) \delta^3({\bf r})- \nonumber
\\
& & - {\sigma \over 6} ( {1\over m_1^2} + {1\over m_2^2} -{1\over m_1
m_2} ) \{ {\bf q}^2_{\rm T} r \}_{\rm W}- \nonumber \\
& & - {\sigma \over 2} ( {\sigma_1 \over m_1^2 } +{\sigma_2 \over
m_2^2} ) \cdot ( {{\bf r} \over r} \times {\bf q} ) - {\sigma i
\over 2} [ {1\over m_1}  {\alpha_1 \cdot {\bf r} \over r} - {1\over
m_2} { \alpha_2 \cdot {\bf r} \over r} ],
\end{eqnarray}
where now ${\bf q}$ stands for the momentum operator and
 the symbol $\{ \,\,\}_{\rm
W}$ stands for the Weyl ordering
prescription.
Notice the nonhermitian  terms in
the Dirac matrices  ${\bf \alpha}_1$ and $ {\bf
\alpha_2}$. Such  terms can be eliminated
 by  performing the Foldy--Wouthuysen  transformation with the
nonhermitian
 generator
\begin{equation}
S= {i \over 2 m_1} \alpha_1 \cdot {\bf q} - {i\over 2 m_2}
 \alpha_2 \cdot {\bf q}
\end{equation}
and we end up with the semirelativistic  potential
\begin{eqnarray}
V && = -  \frac{4}{3}
 \frac{{\alpha}_s}{r} + \sigma r-
  \frac{1}{2m_1m_2} \left\{
 \frac{4}{3} \frac{{\alpha}_s}{r}
(\delta^{hk} + \hat{r}^h \hat{r}^k) q^h q^k \right\}_{{\rm W}} -
\nonumber\\
{} & - & \sum_{j=1}^2 \frac{1}{6m_j^2} \{ \sigma \, r \,
 {\bf q}_{{\rm T}}^2  \}_{{\rm W}} +
\frac{1}{6m_1m_2} \{ \sigma \, r \,
{\bf q}_{{\rm T}}^2  \}_{{\rm W}}+
\nonumber \\
& & +
\frac{1}{8} \left( \frac{1}{m_1^2}
 + \frac{1}{m_2^2} \right)
\nabla^2 \left( - \frac{4}{3} \frac{\alpha_s}{r} + \sigma r \right)
+
\nonumber\\
{} &+&  \frac{1}{2} \left(
 \frac{4}{3} \frac{\alpha_s}{r^3} -
\frac{\sigma}{r} \right) \left[ \frac{1}{m_1^2} {\bf S}_1 \cdot
( {\bf r} \times {\bf q} ) + \frac{1}{m_2^2} {\bf S}_2 \cdot
( {\bf r} \times {\bf q} ) \right] +
\nonumber\\
{} &+& \frac{1}{m_1m_2} \frac{4}{3} \frac{\alpha_s}{r^3} [ {\bf S}_2
\cdot ( {\bf r} \times {\bf q} ) + {\bf S}_1 \cdot ( {\bf r}
\times {\bf q} )] +
\nonumber\\
&+& \frac{1}{m_1m_2} \frac{4}{3} \alpha_s \left\{ \frac{1}{r^3}
\left[ \frac{3}{r^2} ({\bf S}_1 \cdot {\bf r})({\bf S}_2 \cdot
{\bf r}) - {\bf S}_1 \cdot {\bf S}_2 \right] +
\frac{8\pi}{3} \delta^3({\bf r}) {\bf S}_1 \cdot {\bf S}_2 \right\}
\>
\end{eqnarray}
 which coincides with the Wilson loop
 one \cite{bp}.

 \end{document}